\documentclass[letterpaper,twocolumn,prl,aps,superscriptaddress,amsmath,amssymb,floatfix]{revtex4-1}
\usepackage{mathptmx}
\usepackage{ulem}
\usepackage[latin9]{inputenc}
\usepackage{color}
\usepackage{verbatim}
\usepackage{float}
\usepackage{amsmath}
\usepackage{amssymb}
\usepackage{graphicx}
\usepackage{esint}
\usepackage{multirow}
\usepackage[unicode=true,
bookmarks=true,bookmarksnumbered=false,bookmarksopen=false,
breaklinks=false,pdfborder={0 0 1},backref=false,colorlinks=true]
{hyperref}
\hypersetup{
	linkcolor=magenta,urlcolor=blue,citecolor=blue,pdfstartview={FitH},hyperfootnotes=false}

\makeatletter

\usepackage{textcomp}
\usepackage{epstopdf}

\pdfpageheight\paperheight
\pdfpagewidth\paperwidth

\@ifundefined{textcolor}{}{%
	\definecolor{BLACK}{gray}{0}
	\definecolor{WHITE}{gray}{1}
	\definecolor{RED}{rgb}{1,0,0}
	\definecolor{GREEN}{rgb}{0,1,0}
	\definecolor{BLUE}{rgb}{0,0,1}
	\definecolor{CYAN}{cmyk}{1,0,0,0}
	\definecolor{MAGENTA}{cmyk}{0,1,0,0}
	\definecolor{YELLOW}{cmyk}{0,0,1,0}
}

\usepackage{xcolor}
\usepackage{soul}
\newcommand{\ket}[1]{\ensuremath{\left|#1\right\rangle}}

\definecolor{blue}{rgb}{0,0,1}
\definecolor{red}{rgb}{0,0,0}
\definecolor{green}{rgb}{0,1,0}
\newcommand{\red}[1]{\textcolor{red}{ #1}}

\usepackage{soul}

\makeatother

\begin{document}
	\title{Quantum state transfer between superconducting cavities via exchange-free interactions}
	\author{Jie Zhou}
	\affiliation{Center for Quantum Information, Institute for Interdisciplinary Information Sciences, Tsinghua University, Beijing 100084, China}

	\author{Ming Li}
	\affiliation{CAS Center For Excellence in Quantum Information and Quantum Physics,
		University of Science and Technology of China, Hefei, Anhui 230026,
		P. R. China.}
	\author{Weiting Wang}
	\email{wangwt2020@mail.tsinghua.edu.cn}
	\affiliation{Center for Quantum Information, Institute for Interdisciplinary Information Sciences, Tsinghua University, Beijing 100084, China}
	
	\author{Weizhou Cai}
	\affiliation{CAS Center For Excellence in Quantum Information and Quantum Physics,
		University of Science and Technology of China, Hefei, Anhui 230026,
		P. R. China.}
	\author{Ziyue Hua}
	\affiliation{Center for Quantum Information, Institute for Interdisciplinary Information Sciences, Tsinghua University, Beijing 100084, China}
	\author{Yifang Xu}
	\affiliation{Center for Quantum Information, Institute for Interdisciplinary Information Sciences, Tsinghua University, Beijing 100084, China}
	\author{Xiaoxuan Pan}
	\affiliation{Center for Quantum Information, Institute for Interdisciplinary Information Sciences, Tsinghua University, Beijing 100084, China}
	\author{Guangming Xue}
	\affiliation{Beijing Academy of Quantum Information Sciences, Beijing, China}
	\affiliation{Hefei National Laboratory, Hefei, China}
	
	\author{Hongyi Zhang}
	\affiliation{Center for Quantum Information, Institute for Interdisciplinary Information Sciences, Tsinghua University, Beijing 100084, China}
	\author{Yipu Song}
	\affiliation{Center for Quantum Information, Institute for Interdisciplinary Information Sciences, Tsinghua University, Beijing 100084, China}
	
	\author{Haifeng Yu}
	\affiliation{Beijing Academy of Quantum Information Sciences, Beijing, China}
	\affiliation{Hefei National Laboratory, Hefei, China}
	\author{Chang-Ling Zou}
	\email{clzou321@ustc.edu.cn}
	\affiliation{CAS Center For Excellence in Quantum Information and Quantum Physics,
		University of Science and Technology of China, Hefei, Anhui 230026,
		P. R. China.}
	\affiliation{Hefei National Laboratory, Hefei, China}
	
	\author{Luyan Sun}
	\email{luyansun@tsinghua.edu.cn}
	\affiliation{Center for Quantum Information, Institute for Interdisciplinary Information Sciences, Tsinghua University, Beijing 100084, China}
	\affiliation{Hefei National Laboratory, Hefei, China}
	
	\begin{abstract}
		We propose and experimentally demonstrate a novel protocol for transferring quantum states between superconducting cavities using only continuous two-mode squeezing interactions, without {exchange} of photonic excitations between cavities. This approach conceptually resembles quantum teleportation, where quantum information is transferred between different nodes without directly transmitting carrier photons.  In contrast to the discrete operations of entanglement and Bell-state measurement in teleportation, our scheme is symmetric and continuous. We experimentally realize coherent and bidirectional transfer of arbitrary quantum states, including bosonic quantum error correction codes. Our results offer new insights into the quantum state transfer and quantum teleportation. In particular, our demonstration validates a new approach to realize quantum transducers, and might find applications in a wide range of physical platforms.
	\end{abstract}
	
	\maketitle
	
	\textit{Introduction.-} Quantum state transfer (QST), the ability to faithfully transmit quantum information between different nodes in a quantum network, is a fundamental requirement for various quantum information processing tasks, including quantum communication~\cite{BENNETT20147,PhysRevA.59.1829}, distributed quantum computing~\cite{PhysRevLett.78.3221,Kimble2008,PhysRevA.76.062323}, and quantum cryptography~\cite{RevModPhys.74.145,Yin2020}. Conventionally, the transfer of classical information is realized by directly transmitting physical carriers through exchange interactions, such as propagating modes of microwave, acoustic, or optical photons, or through the transduction between different types of information carriers. Such an approach has been generalized to a single quanta level, and the QST based on exchange interaction has been realized in various physical systems including superconducting circuits~\cite{PRXQuantum.2.030321,RN1,RN3,PhysRevLett.120.200501,Leung2019,Zhong2021}, trapped ions~\cite{RevModPhys.82.1209,RN5,feng2023realization}, and neutral atoms \cite{RN2,PhysRevLett.119.010402,Moehring2007}. However, this concept is challenged in the quantum domain by the protocol of quantum teleportation~\cite{PhysRevLett.70.1895}, which allows the transfer of an arbitrary state without sending quantum information carriers.

	
	Quantum teleportation~\cite{PhysRevLett.70.1895,RN6,Hu2023} relies on pre-shared entangled states between the sender (quantum information carrier A) and the receiver (carrier B), followed by a local Bell state measurement between carrier A and an unknown quantum state on carrier C on the sender's side, and classical communications to the receiver. In this sense, quantum information is disembodied from the carrier and transferred. Previously, quantum teleportation has been studied and generalized within a quantum circuit framework, involving discrete gate operations on different nodes. Inspired by the concept of quantum teleportation, an interesting question arises: could the idea of carrier-exchange-free QST be generalized to a continuous evolution system by continuously generating and detecting quantum entanglement between nodes?
	
	\begin{figure}[t!]
		\centering
		\includegraphics{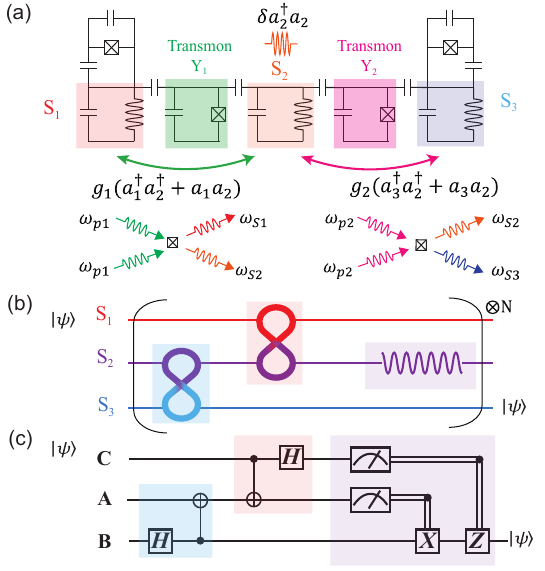}
		\caption{Principle of the exchange-free quantum state transfer (QST). (a) Effective circuit model of the experimental system and the concept of the experiment. (b) Trotterization of the QST process. (c) The quantum circuit for standard quantum teleportation protocol. {The similarity between (b) and (c) is notable.}}
		\label{fig:fig1}
	\end{figure}
	
	
	In this Letter, we address this question by proposing and experimentally demonstrating a novel scheme for QST between superconducting cavities using only continuous two-mode squeezing interactions, without any physical process that exchanges photon excitations. In contrast to the unidirectional feature of quantum teleportation, our scheme is symmetric and enables coherent and bidirectional QST. We realize a counter-intuitive coherent swap operation between two cavities, and the coherent and bidirectional nature is verified by Hong-Ou-Mandel (HOM) interference experiments~\cite{RN108,PhysRevX.8.021073}. Furthermore, the scheme is applicable to arbitrary quantum states in bosonic modes, as demonstrated by the QST of error-correctable bosonic codes~\cite{RN22,RN21,RN109} with the photon-loss error during the QST process being mitigated. Our scheme is universal for bosonic systems and can be extended to phononic modes, trapped ions, and hybrid quantum systems. 
	\red{Moreover, our scheme extends the toolbox of coherent quantum transduction and frequency conversion, offering more flexibility in QST. This approach has the potential to address practical challenges and limitations in various application scenarios, such as more efficient microwave-optical frequency transduction~\cite{Jiang2020}, coherent conversion between telecommunication channels, and robust QST in superconducting circuits (see Ref.~\cite{supplement} for details).}
	
	
	
	\textit{Principle.-} {Figure~\ref{fig:fig1}(a) depicts the effective circuit model of the circuit quantum electrodynamics (QED) system~\cite{RN29}, consisting of three microwave cavities ($S_{1}$, $S_{2}$, and ${S_3}$) in a chain and two auxiliary transmon qubits ($Y_1$ and $Y_2$) to mediate the neighboring cavities. The goal is to transfer quantum states between the two end nodes of the chain.} All the cavities have distinct mode frequencies, and thus the excitation exchange between adjacent nodes are significantly suppressed and there is no direct {interface} between $S_{1}$ and ${S_3}$. Since a driven transmon qubit can stimulate parametric couplings through a four-wave mixing process~\cite{PhysRevX.8.021073,PhysRevA.99.012314}, a two-mode squeezing interaction ($\hbar=1$)
	\begin{equation}
		H_\mathrm{TMS_{1(2)}}=g_{1(2)}(\hat a^\dagger_{1(3)} \hat a^\dagger_{2}+\hat a_{1(3)} \hat a_{2})
	\end{equation}
	between ${S_{1(3)}}$ and ${S_2}$ can be implemented by driving the shared transmon qubit $Y_{1(2)}$. Here,  $\hat a_j$ denotes the annihilation operator of cavity $S_j$ and $g_{1(2)}$ is the induced coupling strength proportional to the drive amplitude. The two drive frequencies are ${\omega_{p1(2)}=(\omega_{S1(3)}+\omega_{S2}+\delta)/2}$  with a common detuning $\delta$ and {$\omega_{Sj}$ being the frequency of cavity $S_j$}, giving an additional Hamiltonian term $H_\mathrm{detune}=\delta\hat a^\dagger_{2} \hat a_{2}$. 
	
	From the view of a continuously evolving quantum system, the detuned $S_2$ with $\delta\gg g_{1,2}$ can be adiabatically eliminated and we arrive at an effective BS Hamiltonian 
	\begin{equation} 
		\hat H_{\mathrm{eff}}=g_\mathrm{eff}(\hat a^\dagger_{1} \hat a_{3}+\hat a_{1} \hat a^\dagger_{3}),
		\label{eq:BS}
	\end{equation}
	although the original Hamiltonian $H_\mathrm{TMS_{1,2}}$ does not contain any excitation exchange terms. The effective Hamiltonian indicates a faithful QST between modes $S_{1}$ and $S_{3}$. \red{In principle, the process of our protocol is similar to that of the Raman method~\cite{RevModPhys.70.1003,Mundhada_2017}. However, it is important to note that our derivation and process focus on three modes, rather than on three energy levels.}
	
	The remarkable coherent exchange coupling through the excitation exchange-free interaction can be interpreted by comparing the analogous quantum circuits of our scheme [Fig.~\ref{fig:fig1}(b)] with the standard quantum teleportation [Fig.~\ref{fig:fig1}(c)]. In our scheme, the continuous time evolution of the system $\hat U(t)=e^{-i(\hat H_\mathrm{TMS1}+\hat H_\mathrm{TMS2}+\hat H_\mathrm{detune})t}$ can be discretized into a repetition of Trotter steps as $\hat U(\Delta t)\approx e^{-i\hat H_\mathrm{TMS1}\Delta t}e^{-i\hat H_\mathrm{TMS2}\Delta t}e^{-i\hat H_\mathrm{detune}\Delta t}$, by choosing a sufficiently small time interval $\Delta t\ll1$. 
	{In comparison to Fig.~\ref{fig:fig1}(c), the two-mode squeezing interactions resemble the Bell state generation (blue shadow) and detection (red shadow). The fast rotation of mode $S_2$ suppresses its population accumulation during the interactions, and thus disentangles it from the other two modes (in analogy to the feedforward process to disentangle the intermediate modes).} By continuously repeating the circuit in Fig.~\ref{fig:fig1}(b), the quantum state will be gradually transferred from $S_1$ to $S_3$.
	
	
	Consequently, our scheme can be viewed as a continuous version of the quantum teleportation protocol. However, due to the symmetry between $S_1$ and $S_3$, reordering the Bell state generation and detection can lead to a reversed coherent QST from $S_3$ to $S_1$, resulting in a bidirectional QST, i.e., a coherent swap of quantum states between the two modes. 
	In contrast, there are asymmetric feedforward operations and sequential order between the two entanglement operations in the standard quantum teleportation protocol, therefore, the state can only be transferred unidirectionally from {$C$ to $B$} in Fig.~\ref{fig:fig1}(c).
	
	
	
	\begin{figure*}
		\centering
		\includegraphics{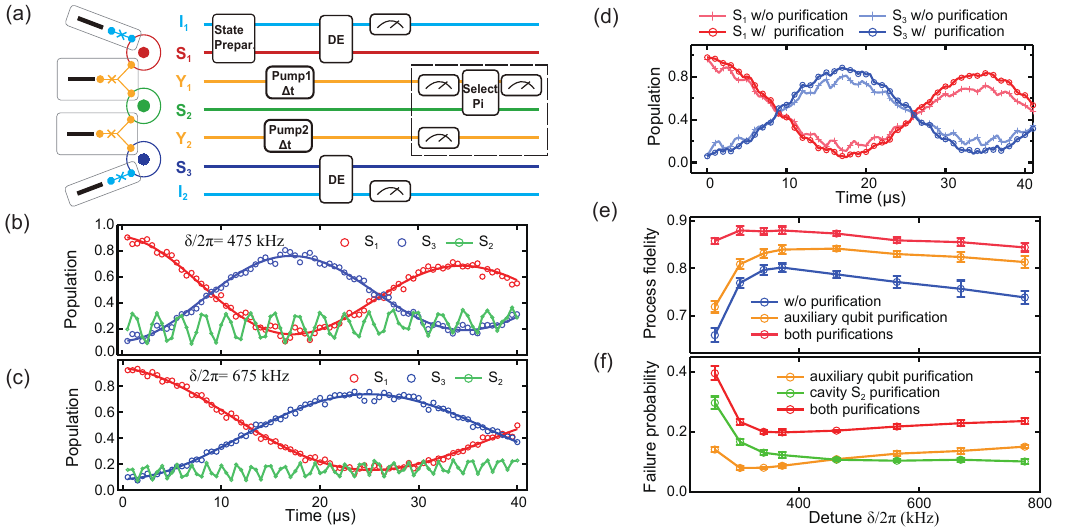}
		\caption{Single-photon QST experiment. (a) Schematic of the circuit-QED system and the experimental sequence. (b) and (c) Population oscillations of $S_1$, $S_2$, and $S_3$ with an initial state $\left| 10 \right\rangle$ for pump detunings $\delta/2\pi= 475$~kHz and $\delta/2\pi= 675$~kHz, respectively. Here, $
			\ket{mn}=\ket{m}_{S_1} \otimes \ket{n}_{S_3}$ is the tensor product Fock state of $S_1$ and $S_3$. The populations of $S_1$ and $S_3$ are fitted to $e^{t/\tau_1}[1+e^{t/\tau_\phi}\mathrm{cos}(\omega t)]$ shown as solid lines. \red{The extracted coherence times are $\tau_1=350~\mu$s and $\tau_\phi=113~\mu$s for (b), while $\tau_1=357~\mu$s and $\tau_\phi=140~\mu$s for (c).}  (d) Population oscillations of $S_1$ and $S_3$ for $\delta/2\pi= 463$~kHz with (dot) and without (cross) purifications. {(e) Process fidelities for different $\delta$, without any purification (blue), with the auxiliary qubit purification (orange), and with both purifications (red). (f) Failure probability for the auxiliary qubit purification (orange), cavity $S_2$ purification (green), and both purifications combined (red).}}
	\label{fig:One_photon}
\end{figure*}

\textit{Single-photon experiments.-} We start with a single-photon experiment, depicted in Fig.~\ref{fig:One_photon}(a). The cavity states are manipulated and detected independently with two {control qubits ${I_1}$ and ${I_2}$}. Initially, Fock state $\left| 1 \right\rangle$ is prepared in cavity ${S_1}$. Subsequently, two pumps are applied to ${Y_1}$ and ${Y_2}$ for a variable time ${t}$, with the pump strength being carefully set to maintain $g/2\pi= g_{1}/2\pi= g_{2}/2\pi\approx 80$~kHz. 
Then, quantum information is decoded from the cavities to the control qubits ${I_1}$ and ${I_2}$ for measurements.
The control pulses during the state preparation and decoding process are optimized using the gradient ascent pulse engineering method~\cite{KHANEJA2005296}. 

Figures~\ref{fig:One_photon}(b) and \ref{fig:One_photon}(c) depict the population evolution of ${S_1}$ and ${S_3}$ with pump detunings {$\delta/2\pi= 475~\mathrm{kHz}$ and $\delta/2\pi=675~\mathrm{kHz}$}, respectively. The populations exhibit opposite oscillations for ${S_1}$ (red) and ${S_3}$ (blue), indicating the transfer of photons between the two cavities. 
By solving the Heisenberg equations \cite{supplement}, we derive the oscillation period $\tau_\mathrm{ST}=\pi/(\frac{\delta}{2}-\sqrt{2}\Omega)$ with $\Omega=\sqrt{\delta^2/8-g^2}$. It is noteworthy that the oscillation occurs only when $\delta>2\sqrt2g$, while a quantum phase transition occurs when $\delta<2\sqrt2g$, causing the cavity populations to diverge. This work will focus solely on the scenario where $\delta>2\sqrt2g$. 
{However, there are unwanted residual photon excitations in $S_2$, which oscillate over time for a given set of initial photon numbers in the three cavities $(n_1, 0, n_3)$ and can be described as $n_{2}(t)=(2+n_1+n_3)(\frac{g}{2\Omega})^2(1-\mathrm{cos}(2\sqrt{2}\Omega t))$.} It reveals that a larger $\delta$ leads to the suppression of $n_{2}$.
However, the QST becomes slower. Such a trend is confirmed by comparing the results in Figs.~\ref{fig:One_photon}(b) and \ref{fig:One_photon}(c). According to the oscillation period of $n_2$ as $\tau_{S2}=\pi/\sqrt{2}\Omega$, there exist sweet points, where $\tau_\mathrm{ST}/\tau_{S2}\in\mathbb{Z}$, such that $n_2$ oscillates to its minimum point when a swap between $S_1$ and $S_3$ is achieved. 


{To mitigate the impact of the residual photons in $S_2$ and the pump-induced excitations in the driven auxiliary qubits $Y_{1}$ and $Y_2$, we perform a two-step purification protocol on $Y_{1}$, $Y_2$, and $S_2$, as indicated by the dashed box in Fig.~\ref{fig:One_photon}(a). Firstly, we purify the two auxiliary qubits $Y_1$ and $Y_2$ after the decoding and the tomography process by only retaining data when both qubits are measured in their ground states, thus effectively eliminating the auxiliary qubit excitations. Subsequently, we further purify cavity $S_2$ by applying a photon-number selective ${\pi}$ pulse to ${Y_1}$~\cite{RN113}, which can only flip ${Y_1}$ when both ${S_1}$ and ${S_2}$ are in the vacuum state. \red{Since the population of $S_1$ has already been swapped to the transmon qubit during the decoding process}, the excitation of $Y_1$ in this step indicates the absence of residual photons in $S_2$, allowing us to retain data that confirms no leakage in $S_2$. This two-step purification protocol enhances the single-photon QST process, as evidenced by the improved visibility of the population oscillations of $S_1$ and $S_3$ in Fig.~\ref{fig:One_photon}(d).}

{There exists a trade-off for the detuning $\delta$, which 
	affects the QST time, to balance the excitation leakage to $S_{2}$ with the pump-induced excitation of the auxiliary qubits and single-photon loss.} In Figs.~\ref{fig:One_photon}(e) and \ref{fig:One_photon}(f), we examine the impact of $\delta$ on the state-transfer efficiency while ensuring the sweet point condition is satisfied. Here, process fidelity is utilized to evaluate the QST performance. 
As shown in Fig.~\ref{fig:One_photon}(e), as $\delta$ increases, the process fidelity without purifications first increases and then decreases. This behavior can be explained as follows. When $\delta$ is small, the dominant error is the large leakage to $S_2$. As $\delta$ increases, the leakage to $S_2$ is suppressed, increasing the fidelity. However, further increases in $\delta$ extend the QST time $\tau_\mathrm{ST}$, causing the accumulation of both single-photon loss and pump-induced auxiliary qubit excitations on $Y_1$ and $Y_2$. Consequently, the QST fidelity gradually decreases for large $\delta$. The impact of 
different errors is also reflected in the change of the failure probability in the purification protocol, as shown in Fig.~\ref{fig:One_photon}(f). With increasing $\delta$, the failure probability of the cavity purification decreases due to the suppression of leakage to $S_2$. Simultaneously, the failure probability of the auxiliary qubit purification increases due to a longer pump time, {except for the first data point (yellow)}. The anomaly in the first point is attributed to the heating of the whole system when $\delta$ approaches $2\sqrt{2}g$.

Figure~\ref{fig:One_photon}(e) shows that the highest process fidelity improves from 80.1\% to 87.9\% after applying purifications. Errors in state preparation and measurement are not excluded here, contributing largely to the infidelity~\cite{supplement}. In our system, the QST fidelity is primarily limited by the strength of the two-mode squeezing interactions. By replacing the auxiliary transmons with other couplers such as SNAILs or flux-pumped SQUIDs~\cite{PRXQuantum.4.020355,Lu_NaturecomRef2}, $g_{1}$ and $g_2$ can be increased by an order of magnitude. {Then, the QST can be implemented more quickly and the photon leakage to $S_2$ can also be further suppressed by utilizing a larger $\delta$.}

\begin{figure}
	\centering
	\includegraphics{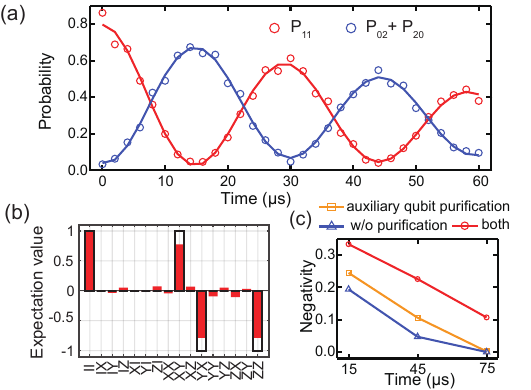}
	\caption{HOM interference Experiment. (a) Population oscillations among $\left| 11 \right\rangle$, $\left| 02 \right\rangle$, and $\left| 20 \right\rangle$ with an initial state $\left| 11 \right\rangle$ and $\delta/2\pi=775$~kHz. The populations of $\left| 02 \right\rangle$ and $\left| 20 \right\rangle$ are added together and shown as blue dots. 
		\red{The extracted coherence times are $\tau_1=121~\mu$s and $\tau_\phi=198~\mu$s.} (b) Reconstructed two-qubit Pauli operator expectation values in the $\{\left| 0 \right\rangle$, $\left| 2 \right\rangle \}$ subspaces of both ${S_1}$ and ${S_3}$ with a pump time of $\Delta t=$15 ${\mu}$s. 
		(c) Negativity of the reconstructed two-qubit state for $\Delta t=$ 15~${\mu}$s, 45~${\mu}$s, and 75~${\mu}$s.}
	\label{fig:HOM}
\end{figure}
\textit{HOM interference.-}	Different from the standard quantum teleportation protocol, our QST is symmetric to nodes $S_{1}$ and $S_{3}$, as indicated by the bidirectional BS-type interaction (Eq.~\ref{eq:BS}), and such a feature can be proved by the HOM interference between ${S_1}$ and ${S_3}$ \cite{RN108,PhysRevX.8.021073}. By initializing the two cavities to a joint photon number state $\left| 11 \right\rangle $, and we monitor the population evolution of the photon number states $\left| 11 \right\rangle$, $\left| 02 \right\rangle$, and $\left| 20 \right\rangle$ after turning on the interaction. Here, $|mn\rangle=|m\rangle_{S_{1}} \otimes |n\rangle_{S_{3}}$ is the tensor product Fock state of $S_1$ and $S_3$. The populations of different joint number states are obtained by applying photon-number selective $\pi$ pulses on {the control qubits $I_1$ and $I_2$}, with the population $P_{mn}$ given by the joint probability that both $I_1$ and $I_2$ are simultaneously flipped. 

Figure~\ref{fig:HOM}(a) shows the results for the pump detuning $\delta/2\pi= 775$~kHz, demonstrating clear oscillation between $P_{11}$ and $P_{02}+P_{20}$. At half the period time $\tau_\mathrm{ST}/2$, $P_{02}+P_{20}$ reaches its maximum, highlighting the bunching effect of bosons. Here, only the auxiliary qubit purification can be performed because $S_1$ is not reset to the vacuum. To further investigate the coherence between $|20\rangle $ and $|02\rangle $ and the entanglement between $S_{1}$ and $S_{3}$, we decode the quantum information from $\left\{|0\rangle,|2\rangle \right\}$ subspaces of the cavities to the control qubits $I_1$ and $I_2$ for a joint state tomography. Figure~\ref{fig:HOM}(b) shows the reconstructed two-qubit Pauli operator expectation values with an interaction time $\tau_\mathrm{ST}/2\approx 15~\mu$s (red bar). The data is obtained after the two-step purification. The ideal results for state $(\left| 02 \right\rangle+i\left| 20 \right\rangle)/\sqrt2$ are also shown as black frames for comparison, resulting in a state fidelity of $81\%$. For longer BS interaction times $(n+1)\tau_\mathrm{ST}/2$ with $n\in \mathbb{Z}$, we also get the density matrices and calculate their {negativity~\cite{Vidal2002PRA}} to justify the quantum entanglement between $S_{1}$ and $S_{3}$, as shown in Fig.~\ref{fig:HOM}(c). Without purification, the negativity is still larger than zero at $45~\mu$s, indicating the existence of entanglement. After the two-step purification, entanglement can be maintained for a longer time of 75~$\mu$s.

\begin{figure}
	\centering
	\includegraphics{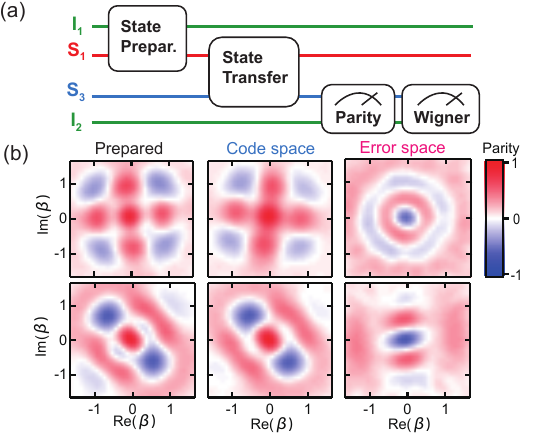}
	\caption{QST for different states with a pump detuning $\delta/2\pi= 463~\mathrm{kHz}$. (a) Experimental sequence for QST of error-correctable logical states. (b) Prepared and received Wigner functions for binomial logical states $\left| 0_L \right\rangle$ and $\left|+i_L \right\rangle$.} 
\label{fig:Wigner}
\end{figure}

\textit{QST of bosonic logical codes.-} The coherent BS-type interaction between bosonic modes in Eq.~\ref{eq:BS} enables QST to work for the entire Hilbert space of the two cavities $S_1$ and $S_3$, including both multi-photon and superposition states. {This capability is crucial for transferring error-correctable logical states of bosonic quantum error correction codes~\cite{RN23, RN109,RN110,RN111,Gertler2021,Campagne2020}, such as a binomial code in our demonstration~\cite{RN22,RN21,RN109}, providing a means to overcome single-photon-loss errors during the transfer process~\cite{PRXQuantum.2.030321}.} 
Figure~\ref{fig:Wigner} depicts the sequence and the results for the transfer of binomial logical states $\left| 0_L \right\rangle=(\left| 0 \right\rangle+\left| 4 \right\rangle)/\sqrt{2}$ and $\left| +i_L \right\rangle=(|0_L\rangle+i|1_L\rangle)/\sqrt{2} =(\frac{\left| 0 \right\rangle+\left| 4 \right\rangle}{\sqrt{2}}+i \left| 2 \right\rangle)/\sqrt{2}$, where $\ket{1_L}=\ket{2}$. We first prepare the logical states in ${S_1}$, and then measure the Wigner function of the states in ${S_3}$ after the QST is accomplished. 
The average state fidelity of the prepared states in ${S_1}$, as characterized also through a Wigner tomography, is found to be $89.6\%$ [left column of Fig.~\ref{fig:Wigner}(b)]. By introducing parity measurements before the final Wigner tomography after the QST, the single-photon-loss errors of the cavities can be detected. An even parity indicates that the states are likely to remain in the code space, and the corresponding transferred states have an average state fidelity of $84.9\%$ [middle column of Fig.~\ref{fig:Wigner}(b)]. On the other hand, an odd parity indicates an error during the transfer process, resulting in the states $\left| 0_E \right\rangle=\left| 3 \right\rangle$ and $\left| +i_E \right\rangle=(\left| 1 \right\rangle+i \left| 3 \right\rangle)/\sqrt{2}$ [righ column of Fig.~\ref{fig:Wigner}(b)]. The coherence is clearly preserved in the error space, allowing the erroneous logical states to be corrected to the code space.

\textit{Conclusion.-} We have experimentally realized QST between two superconducting cavities without any physical process that exchanges photon excitations. The process is constructed by only continuous two-mode squeezing interactions, which can be understood as a continuous analogue of the quantum teleportation protocol. Additionally, the continuous process is symmetric and allows coherent and bidirectional QST, as verified by the HOM experiment. The scheme works for arbitrary states in bosonic modes and the transfer of error-correctable codes is further demonstrated. Our results not only provide a new perspective on QST and quantum teleportation, but also offer novel insights into the realization of quantum networks and distributed quantum computing architectures. Specially, our scheme can be applied in certain experimental scenarios where the direct quantum frequency conversion is difficult but the two-mode squeezing is more efficient~\cite{supplement}.

\smallskip{}

\begin{acknowledgments}
This work was funded by the National Natural Science Foundation of China (Grants No. 11925404, 92165209, 92365301, 12061131011, 92265210, 11890704, 92365206),  Innovation Program for Quantum Science and Technology (Grant No.~2021ZD0300203 and 2021ZD0301800), and the National Key R\&D Program (2017YFA0304303). This work was also supported by the Fundamental Research Funds for the Central Universities and USTC Research Funds of the Double First-Class Initiative. This work was partially carried out at the USTC Center for Micro and Nanoscale Research and Fabrication.

\end{acknowledgments}

\bibliography{ref}

\global\long\def\figurename{ {Supplementary Figure}}%
\renewcommand{\thefigure}{S\arabic{figure}}
\setcounter{figure}{0}
\global\long\def\thepage{S\arabic{page}}%
\setcounter{page}{1}
\global\long\def\theequation{S\arabic{equation}}%
\setcounter{equation}{0}
\global\long\def\tablename{\textbf{Supplementary Table}}%

\begin{center}
	\textbf{{Supplementary Materials}}
	\par\end{center}	
	
	\section{Experimental Setup and System Parameters}
	Our experimental device is composed of three high-Q superconducting cavities ($S_{1}$, $S_{2}$, and ${S_3}$) \cite{PhysRevB.94.014506Ref5} and four transmon qubits ($I_{1}$, $I_{2}$, $Y_{1}$, and ${Y_2}$). The cavity modes $S_{1(3)}$ and $S_{2}$ are coupled together via the auxiliary Y-type transmon $Y_{1(2)}$. $S_{1(3)}$ is also coupled to another I-type control qubit $I_{1(2)}$. Each of the four qubits is coupled with an individual low-Q resonator for fast readout. More details about the fabrication of the device can be found in Ref.~\cite{cai2023protectingRef4}.
	
\begin{figure*}
		\centering
		\includegraphics{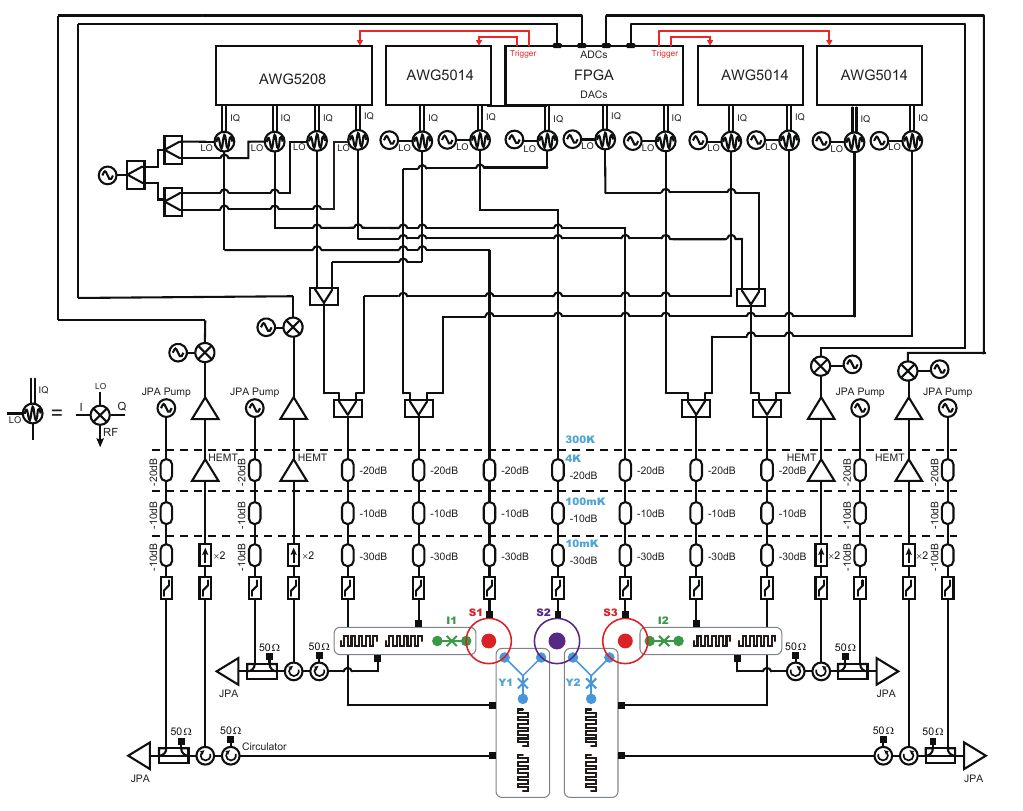}
		\caption[The system wiring]{\textbf{The system wiring.} 
			All the control signals are generated by IQ mixing process. The controls of $S_1$, $S_3$, and the four-wave-mixing pulses share the same local oscillators for phase locking.
		}
		\label{fig:SM_Wiring}
	\end{figure*}

	The experimental setup is depicted in Fig.~\ref{fig:SM_Wiring}. All the control pulses are generated by the IQ mixing process, and the IQ signals are produced by AWGs or the DAC outputs of the field programmable gate array (FPGA) boards. To ensure the phase locking, the drives for $S_1$, $S_3$, and the four-wave-mixing pulses share the same local oscillators. The output signals are amplified by Josephson parametric amplifiers (JPA) \cite{RN9Ref6,PhysRevB.79.184301Ref7}, together with high-electron-mobility-transistor amplifiers at 4 K.
	
	Table ~\ref{tab:SM_Hamiltonian} shows the measured nonlinear coupling strength between different modes. As claimed in the main text, the direct coupling strength between $S_{1}$ and $S_{3}$ is too small to be measured and can be neglected. The coherence parameters of the quantum system are also characterized and shown in Table ~\ref{tab:SM_coherence}.
	\begin{table}[b]
		\centering
		\caption{The measured nonlinear interaction ($ \chi_{ij}/2 \pi $ in MHz) of the system (--- means the term is not measurable; $ \sim $ means the term is not measured).}
		\label{tab:SM_Hamiltonian}
		\resizebox{\linewidth}{!}{
			\begin{tabular}{c|ccccccc} 
				\hline
				Mode   & $I_1$  & $S_1$    & $Y_1$  & $S_2$    & $Y_2$             & $S_3$    & $I_2$      \\ 
				\hline
				$I_1$  & $\sim$ & 1.300    & 0.003  & ---      & 0.005             & ---      & ---        \\
				$S_1$  & 1.300  & 0.003    & 0.623  & $<0.001$ & ---               & ---      & ---        \\
				$Y_1$  & 0.003  & 0.623    & $\sim$ & 1.309    & 0.019             & ---      & ---        \\
				$S_2$  & ---    & $<0.001$ & 1.309  & 0.004    & 0.661             & $<0.001$ & ---        \\
				$Y_2$  & 0.005  & ---      & 0.019  & 0.661    & $\sim$            & 0.451    & 0.027      \\
				$S_3$  & ---    & ---      & ---    & $<0.001$ & 0.451             & 0.002    & 1.411      \\
				$I_2$  & ---    & ---      & ---    & ---      & 0.027             & 1.411    & $\sim$     \\
				\hline
			\end{tabular}
		}
	\end{table}
	
	\begin{table}
		\centering
		\caption{The measured resonant mode frequencies, coherence times, and thermal populations.}
		\label{tab:SM_coherence}
		\begin{tabular}{c|cccc} 
			\hline
			Mode          & Frequency (GHz) & $T_1$ ($\mu$s)  & $T_2^*$ ($\mu$s) & ${n_\mathrm{th}}$ \\ 
			\hline
			$I_1$         & 4.209           & 80-110                               & 30-80                             & 4$\%$       \\
			$S_1$         & 6.102           & 265                               & ---                               & 3$\%$ 			\\
			$Y_1$         & 4.816           & 70                                & 20-50                                & 2$\%$       \\
			$S_2$         & 5.560           & 300                               & ---                               & 2$\%$      \\
			$Y_2$         & 4.461           & 110-150                           & 20-60                               & 2.5$\%$  \\
			$S_3$         & 6.005           & 314                               & ---                               & 2.5$\%$      \\
			$I_2$         & 4.220           & 50                            & 30-80                             & 4$\%$   \\
			\hline
		\end{tabular}
	\end{table}
	\section{The effective Hamiltonian}
	\label{sec:effe}
	\subsection{Two-mode squeezing interaction}
	The two-mode squeezing (TMS) interaction between $S_{1(3)}$ and $S_{2}$ is realized by applying an off-resonance pump tone to the auxiliary transmon qubit $Y_{1(2)}$. {Here, we only derive this interaction between $S_1$ and $S_2$ with the one for $S_2$ and $S_3$ being similar.} The full Hamiltonian with the external pump can be described as $(\hbar=1)$ 
	\begin{equation} 
		\begin{aligned}
			H_{} =&\omega_{s_{1}} \hat a^\dagger_{1}\hat a_{1}+\omega_{s_2} \hat a^\dagger_{2}\hat a_2+\omega_{q} \hat q^\dagger \hat  q\\&-E_J(\cos\hat{\varphi}+\hat\varphi^2/2)+2\mathrm{Re}(\epsilon_p e^{-i\omega_p t})(\hat q^\dagger +\hat q),
		\end{aligned}
		\label{eq:initial Hamiltonian}
	\end{equation}
	with
	\begin{equation*}
		\hat \varphi=\phi_{s_1}(\hat a_1^\dagger +\hat a_1)+\phi_{s_2}(\hat a_2^\dagger +\hat a_2)+\phi_{q}(\hat q^\dagger +\hat q).					
	\end{equation*}
	{Here, $\hat a_j$ is the annihilation operator for cavity $S_j$ with a mode frequency of $\omega_{s_j}$, $\hat q$ is the annihilation operator for the qubit mode with a frequency of $\omega_q$, $E_J$ is the Josephson energy of the qubit, $\omega_p$ is the frequency of the pump with an amplitude of $\epsilon_p$, and $\phi_k$ is the zero point fluctuation of the corresponding mode $k$.} 
	
	In our experiment, $\omega_p$ satisfies the condition
	\begin{equation}
		2\omega_{p}=\omega_{s_{1}}+\omega_{s_2}.
		\label{eq:freq}
	\end{equation}
	Similar to the treatment in Ref.~\cite{PhysRevX.8.021073Ref8}, we perform the following unitary transformation to eliminate the fast rotation terms:
	\begin{equation} 
		\hat U=e^{-i\omega_qt\hat q^\dagger \hat  q}e^{-i\omega_{s_1}t\hat a^\dagger_{1}\hat a_1}e^{-i\omega_{s_2}t\hat a^\dagger_{2}\hat a_2}.
		\label{eq:unitary}
	\end{equation}
	Then, we utilize a displacement transformation to eliminate the pump term, and $\hat q$ becomes:
	\begin{equation} 
		\begin{aligned}
			&\hat q \rightarrow \hat q +\xi e^{-i\omega_p\hat q^\dagger \hat  q},\\
			&\xi=-\frac{i\epsilon_d}{\kappa/2+i(\omega_q-\omega_d)},
		\end{aligned}
		\label{eq:displace}
	\end{equation}
	where $\kappa$ is the decay rate of the qubit. Then, we can get the effective Hamiltonian by expanding the cosine term in Eq. ~\ref{eq:initial Hamiltonian} to the 4th order and keeping only the non-rotating terms:
	\begin{equation}
		\begin{aligned}
			\hat H=& \hat H_\mathrm{TMS}+\hat H_\mathrm{SS},\\
			\hat H_\mathrm{TMS}=&-E_J \phi_q^2\phi_{s_1}\phi_{s_2}(\xi^2 \hat a^\dagger_{1} \hat a^\dagger_{2}+(\xi^*)^2 \hat a_{1} \hat a_{2}),\\
			\hat H_\mathrm{SS}=&-E_J \phi_q^4|\xi|^2\hat q^\dagger \hat  q .\\
		\end{aligned}
	\end{equation}
	Here, $\hat H_\mathrm{TMS}$ is the TMS interaction that we want and $\hat H_\mathrm{SS}$ is the Stark shift term caused by the pump. Note that only the terms including $\xi$ are shown here.
	
	\subsection{State transfer Hamiltonian}
	Applying drive tones satisfying Eq.~\ref{eq:freq} to the auxiliary transmons $Y_1$ and $Y_2$, we can get the desired TMS interactions. However, to realize the state transfer between $S_1$ and $S_3$, we need to decouple $S_2$ from both $S_1$ and $S_3$. We achieve this in the experiment by adding a common detuning $\delta$ to the pump tones:
	\begin{equation}
		2\omega_{p_{1(2)}}-\delta=\omega_{s_{1(3)}}+\omega_{s_2}.
	\end{equation}
	Then, we can get a Hamiltonian as:
	\begin{equation}
		\hat H_{}= g_1(\hat a^\dagger_{1} \hat a^\dagger_{2}+\hat a_{1} \hat a_{2})+g_2(\hat a^\dagger_{3} \hat a^\dagger_{2}+\hat a_{3} \hat a_{2})+\delta\hat a^\dagger_{2} \hat a_{2},
		\label{eq:Hdetune}
	\end{equation}
	{where $g_1$ and $g_2$ are the induced coupling strengths between the neighboring cavities.} In the limit of $g_1,g_2 \ll \delta$,  $S_2$ can be eliminated from the Hamiltonian:
	\begin{equation}
		\begin{aligned}
			&\hat H_{\mathrm{eff}}= g_{\mathrm{eff}}(\hat a^\dagger_{1} \hat a_{3}+\hat a_{1} \hat a^\dagger_{3}),\\
		\end{aligned}
		\label{eq:Hbs}
	\end{equation} 	
	with $g_{\mathrm{eff}}=g_{1}g_{2}/\delta$.
	
	When $g_1,g_2 \ll \delta$ is not satisfied strictly, the dynamics of the Hamiltonian in Eq.~\ref{eq:Hdetune} can be found by solving the Heisenberg equations. For simplification, we set $g_1=g_2=g$ in the following derivation:
	\begin{equation}
		\begin{aligned}
			\frac{d\hat a_1}{dt} =& -ig \hat a^\dagger_2,\\
			\frac{d\hat a_3}{dt}=& -ig \hat a^\dagger_2,\\
			\frac{d\hat a^\dagger_2}{dt}=& ig (\hat a_1+\hat a_3)-i\delta\hat a^\dagger_2.\\
		\end{aligned}
	\end{equation}
	The three operators $a_{1}$, $a_{3}$, $a^{\dagger}_{2}$ commute with each other and form a closed set of differential equations. The solutions can be written in the following form:	
	\begin{equation}
		\begin{aligned}
			\hat a_1=& c_{11}(t)\hat a_1(0)+c_{12}(t)\hat a^\dagger_2(0)+c_{13}(t)\hat a_3(0),\\
			\hat a_2=& c_{21}(t)\hat a^\dagger_1(0)+c_{22}(t)\hat a_2(0)+c_{23}(t)\hat a^\dagger_3(0),\\
			\hat a_3=& c_{31}(t)\hat a_1(0)+c_{32}(t)\hat a^\dagger_2(0)+c_{33}(t)\hat a_3(0),\\
		\end{aligned}
	\end{equation}
	with the time-dependent coefficients:
	\begin{equation}
		\begin{aligned}
			c_{11}(t)=& c_{33}(t)=\frac{1}{2}[1+e^{\frac{i\delta t}{2}}(\cos\sqrt2\Omega t-\frac{i\delta}{\sqrt8\Omega} \sin\sqrt2\Omega t)],\\
			c_{12}(t)=& c_{32}(t)=c_{21}(t)=c_{23}(t)=-\frac{ig}{\sqrt2\Omega}e^{\frac{i\delta t}{2}}\sin\sqrt2\Omega t,\\
			c_{31}(t)=& c_{13}(t)=\frac{1}{2}[e^{\frac{i\delta t}{2}}(\cos\sqrt2\Omega t-\frac{i\delta}{\sqrt8\Omega} \sin\sqrt2\Omega t)-1],\\
			c_{22}(t)=&e^{\frac{i\delta t}{2}}(\cos\sqrt2\Omega t+\frac{i\delta}{\sqrt8\Omega} \sin\sqrt2\Omega t),
		\end{aligned}
	\end{equation}
	where $\Omega=\sqrt{\frac{\delta^2}{8}-g^2}$. The above solutions are valid only when $\delta>2\sqrt2 g$. Once $\delta<2\sqrt2 g$, all three cavities will go into a regime of parametric oscillation, which is not included in our present discussion. 
	
	Considering $\langle \hat n \rangle=\langle \hat a^\dagger \hat a \rangle$ , we can get the time dependence of the average photon numbers of the three cavities:
	\begin{equation}
		\begin{aligned}
			n_1(t)=& n_1(0)|c_{11}(t)|^2+(1+n_2(0)|c_{12}(t)|^2)+n_3(0)|c_{13}(t)|^2,\\
			n_2(t)=& (1+n_1(0))|c_{21}(t)|^2+n_2(0)|c_{22}(t)|^2\\
			&+(1+n_3(0))|c_{23}(t)|^2,\\
			n_3(t)=& n_1(0)|c_{31}(t)|^2+(1+n_2(0)|c_{32}(t)|^2)+n_3(0)|c_{33}(t)|^2.\\
		\end{aligned}
	\end{equation}
	Assuming that $n_2(0)=n_3(0)=0$, we can get
	\begin{equation}
		\begin{aligned}
			n_1(t)=& n_1(0)|c_{11}(t)|^2+|c_{21}(t)|^2,\\
			n_2(t)=& (2+n_1(0))|c_{21}(t)|^2,\\
			n_3(t)=& n_1(0)|c_{31}(t)|^2+|c_{21}(t)|^2,\\
		\end{aligned}
		\label{eq:nofS}
	\end{equation}
	with 
	\begin{equation}
		|c_{21}|^2=(\frac{g}{\sqrt2\Omega})^2(1-\cos(2\sqrt2\Omega t)).
	\end{equation}
	
	{Given that the terms associated with the state transfer are dependent on the initial photon conditions, the relevant terms are specifically $|c_{11}(t)|^2$ and $|c_{31}(t)|^2$. Based on these terms, we have deduced the state transfer time $\tau_\mathrm{ST}=\pi/(\frac{\delta}{2}-\sqrt{2}\Omega)$.}  
	{However, within the expressions for $n_1(t)$ and $n_3(t)$, we also identify an additional term $|c_{21}(t)|^2$, which is not associated with the initial photon number $n_1(0)$. This term corresponds to the photon leakage into $S_2$ and oscillates with a period of $\tau_{S2}=\frac{\pi}{\sqrt2\Omega}$ and an amplitude of $(\frac{g}{\sqrt2\Omega})^2$. Fortunately, in the limit of $g_1,g_2 \ll \delta$, the amplitude of these oscillations diminishes, allowing us to neglect the terms related to $|c_{21}(t)|^2$ and revert to the Hamiltonian in Eq.~\ref{eq:Hbs}.}
	
	To maximize the efficiency of the state transfer process, we choose the state transfer time to make $c_{21}(\tau_\mathrm{ST})=0$, corresponding to $\tau_\mathrm{ST}=k\tau_{S2}, k\in \mathbb{Z}$.
	
	\section{Calibration of the effective Hamiltonian}  
	\subsection{Calibration of TMS strength}
	
	\begin{figure}
		\centering
		\includegraphics{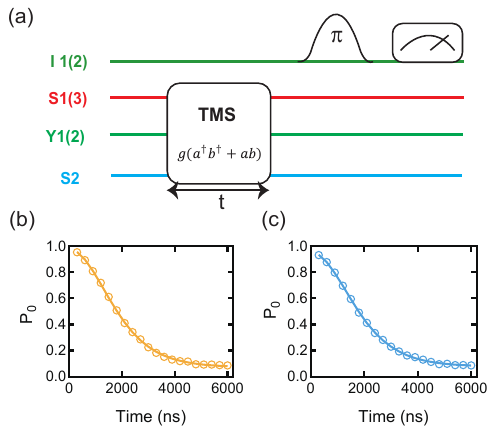}
		\caption[Calibration of the TMS interaction]{\textbf{Calibration of the TMS interactions.} 
			(a) The pulse sequence for the calibration. The system starts from each cavity in the vacuum state and the qubits in their ground state. Subsequently, the interaction is on for a variable time $t$. Then a selective $\pi$ pulse is applied to the control qubit to determine whether cavity $S_{1(3)}$ is in the vacuum state. (b) and (c) The probability of measuring the cavity in the vacuum state $P_0$ varies with the interaction time for systems $S_1-Y_1-S_2$ and $S_3-Y_2-S_2$, respectively. The experimental data points are shown in hollow circles, with the solid lines representing the fits to $P_0=\frac{a}{\cosh^2(gt)}+b$. {The results presented here are obtained when $g_1/2\pi \approx g_2/2\pi \approx 80$~kHz}.
		}
		\label{fig:TMS}
	\end{figure}
	
	Applying the TMS interaction to the vacuum state $\ket{00}$, we create the two-mode squeezed vacuum state, which can be described by
	\begin{equation}
		|\mathrm{TMSV}\rangle=\frac{1}{\cosh r}\sum_{n=0}^{\infty}\tanh^{2n} r|nn\rangle,
	\end{equation}
	where $r=gt$ is the degree of squeezing. The population of the joint Fock state of the two cavities $|nn\rangle$ is then:
	\begin{equation}
		|\langle nn|\mathrm{TMSV}\rangle|^2=(\frac{\tanh^{2n} r}{\cosh r})^2.
		\label{eq:TMSpro}
	\end{equation}
	In our system, we can monitor the population of the Fock state $|n\rangle$ in the cavity by applying a photon-number-selective $\pi$ pulse to the corresponding control qubit~\cite{RN113Ref10}. 
	{As illustrated in Fig. S2, we determine the interaction strength $g_{1(2)}$ by fitting $P_0$, the probability of cavity $S_{1(3)}$ being in the $|0\rangle$ state, to Eq.~\ref{eq:TMSpro}}.
	
	It is worth noting that, due to the AC-Stark shift, the actual frequency for the pump should be equal to $(\omega_{s_{1(3)}}+\omega_{s_2})/2+\Delta_\mathrm{ac}$. Here, $\omega_{s_{1(3)}}$ and $\omega_{s_2}$ denote the frequencies of the cavities in the absence of pumps. If $\Delta_\mathrm{ac}$ is not compensated, it will cause the states in the cavity to oscillate back to $|0\rangle$, resulting in a slower decline in $P_0$. 
	{Therefore, by scanning the pump frequency and locating the point with the steepest decline in $P_0$, we can determine the value of $\Delta_\mathrm{ac}$.}
	
	\subsection{Calibration of detuning $\delta$ }
	When two pumps are applied simultaneously, the Stark shift of $S_2$ will be different from that under a single pump. The equivalent Hamiltonian should be written as:
	\begin{equation}
		\hat H_{}= g_1\hat a^\dagger_{1} \hat a^\dagger_{2}e^{-i(\Delta_0 + \Delta_d) t}+g_2\hat a^\dagger_{3} \hat a^\dagger_{2}e^{-i(\Delta_1 + \Delta_d) t}+h.c.
		\label{eq:Hdetunetwo}
	\end{equation}
	Applying a simple transformation to this Hamiltonian yields:
	\begin{equation}
		\begin{aligned}
			\hat H_{}= &g_1(\hat a^\dagger_{1} \hat a^\dagger_{2}+\hat a_{1} \hat a_{2})+g_2(\hat a^\dagger_{3} \hat a^\dagger_{2}+\hat a_{3} \hat a_{2})\\
			&+(\Delta_0+\Delta_d)\hat a^\dagger_{2} \hat a_{2}+(\Delta_1-\Delta_0)\hat a^\dagger_{1} \hat a_{1}.
			\label{eq:Hdetunetwo2}
		\end{aligned}
	\end{equation}
	where $\Delta_d$ is the detuning added intentionally, $\Delta_0$ and $\Delta_1$ are the detunings caused by the AC-Stark shift. The last term in Eq.~\ref{eq:Hdetunetwo2} will result in incomplete photon exchange when the beam splitter (BS)-type interaction is on. It can be easily eliminated by scanning the frequency of one pump and selecting the position where the oscillation amplitude is maximized. 
	
\begin{figure}
		\centering\includegraphics{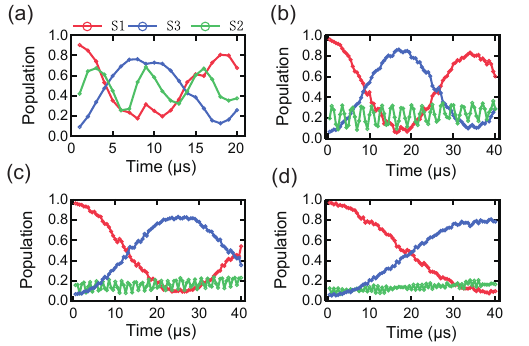}
		\caption[The population oscillation]{\textbf{The population oscillations of $S_1$, $S_2$, and $S_3$ for different $\delta$.} The data presented here for $S_1$ and $S_3$ have undergone the purification process.
			(a), (b), (c), and (d) correspond to $(\Delta_0+\Delta_d)/2\pi=$275~kHz, 475~kHz, 675~kHz, and 975~kHz,
			respectively.
		}
		\label{fig:Different delta}
	\end{figure}

\begin{figure}[h]
		\centering
		\includegraphics{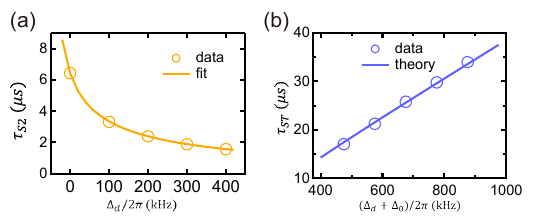}
		\caption[The dependence]{\textbf{The dependence of $\tau_{S2}$ and $\tau_\mathrm{ST}$ on $\delta$.} 
			(a) The dependence of $\tau_{S2}$ on $\Delta_d$. The data is fitted to $\tau_{S2}=\frac{2\pi}{\sqrt{(\Delta_d+\Delta_0)^2-8g^2}}$ to get the value of $\Delta_0$. (b) The dependence of $\tau_\mathrm{ST}$ on $\Delta_d+\Delta_0$. The experimental data align well with the theoretically predicted values, confirming that the obtained $\Delta_0$ through fitting is valid.
		}
		\label{fig:fitdelta0}
	\end{figure}

	However, $\Delta_0$ cannot be obtained directly. Considering that we have already known the strengths of the TMS interactions, we can get the value of $\Delta_0$ from $\tau_{S2}$ or $\tau_\mathrm{ST}$. From Fig.~\ref{fig:Different delta} and Eq.~\ref{eq:nofS},  we observe that when $\delta=\Delta_0+\Delta_d$ is small, the photon oscillations in $S_2$ are more pronounced, while the oscillations between $S_1$ and $S_3$ appear less smooth. When $\Delta_0+\Delta_d$ is large, the oscillations in $S_2$ are suppressed, and simultaneously, the oscillations between $S_1$ and $S_3$ become smoother. To obtain $\Delta_0$, we measure the oscillation period of $S_2$ at positions where the oscillations are more pronounced, and fit it to $\tau_{S2}=\frac{\pi}{\sqrt2\Omega}=\frac{2\pi}{\sqrt{(\Delta_d+\Delta_0)^2-8g^2}}$, as shown in Fig.~\ref{fig:fitdelta0} (a). Through the fitting, we obtain $\Delta_0/2\pi=275$~kHz. To verify the validity of the derived $\Delta_0$, we theoretically calculate $\tau_\mathrm{ST}$ for different $\Delta_d$ and compare the results with the experimental values. The results are shown in Fig.~\ref{fig:fitdelta0}(b) and the experimental data align well with the theoretically predicted values.
	
	\section{Error analysis} 
	
	To further understand the impact of different errors on the results, we analyze the error budget for $\delta/2\pi = 373$ kHz, corresponding to a state transfer time $\tau_\mathrm{ST}=13.3\ \mu$s, which maximizes the process fidelity to 88\%. For the sake of analysis convenience, we assume that the contributions of different errors to the process fidelity are independent. From the process matrix that we obtain in the experiment, we find that the error model can be approximated as a depolarizing channel. We can then express the process fidelity as
	\begin{equation}
		\begin{aligned}
			F_\mathrm{process}=\frac{1}{4}+\frac{3\prod_i P_{i}}{4},
			\label{eq:process}
		\end{aligned}
	\end{equation}
	{where $1-P_i$ represents the contributions of different error types to the infidelity.}	These infidelities can be calculated by comparing the process fidelities that are obtained under different conditions in both experiments and simulations. 
	
	\begin{table}[t]
		\centering
		\caption{The error budget for $\delta/2\pi=373$ kHz.}
		\label{tab:infidelity}
		\begin{tabular}{|cc|c|}
			\hline
			\multicolumn{2}{|c|}{Error type}                                           & Infidelity (1-$P_i$) \\ \hline
			\multicolumn{1}{|c|}{\multirow{3}{*}{quantum state transfer}} & auxiliary qubit excitation          & 7.3 $\%$            \\ \cline{2-3} 
			\multicolumn{1}{|c|}{}                                &  residual photons in $S_2$        & 6 $\%$              \\ \cline{2-3} 
			\multicolumn{1}{|c|}{}                                & cavity decoherence & 4.2 $\%$            \\ \hline
			\multicolumn{2}{|c|}{state preparation + tomography}                            & 8.9 $\%$            \\ \hline
			\multicolumn{2}{|c|}{others}                                               & 3.7 $\%$          \\ \hline
		\end{tabular}
	\end{table}
	
	The results are presented in Table~\ref{tab:infidelity}. The infidelity induced by the auxiliary qubit excitations is obtained by comparing the fidelity without any purification $F_0$ to the fidelity with the auxiliary qubit purification $F_1$, and is calculated to be {$1-(F_0-0.25)/(F_1-0.25)=7.3\%$}. Similarly, the infidelity related to residual photons in $S_2$ is determined {to be 6\%} by comparing the fidelity with the two-step purification to the fidelity with only the auxiliary qubit purification. 
	
	To evaluate the infidelity of the state preparation and tomography process, we perform numerical simulations~\cite{RN112Ref9} using the coherence data in Table~\ref{tab:SM_coherence} with an ideal state transfer process. We get a process fidelity of 93.3\%, corresponding to an infidelity of 8.9\%. Then, we add the cavity decoherence to the state transfer process and obtain a process fidelity of 90.4\%, which results in an infidelity of 4.2\% due to the cavity decoherence. The simulated fidelity (90.4\%) is still slightly higher than the experimental one (88\%). This could be attributed to waveform distortion caused by the experimental equipment and some other errors that are not considered in the simulation, such as the readout error during the purification process. This extra infidelity is about 3.7\%.
	
	\red{Considering that our infidelity is largely due to the long state transfer time, we will next discuss how to shorten the state transfer time without increasing the residual photons in the bus mode. In section~\ref{sec:effe}, we have derived  the expressions of the state transfer time and population oscillations in bus mode $S_2$. Under the assumption that $g \ll \delta$, the state transfer time $\tau_\mathrm{ST}$ is inversely proportional to $g^2/\delta$, while the oscillation amplitude of the bus mode is proportional to $(g/\delta)^2$. In order to lower the swap time without introducing more leakage, we can increase the strength of the TMS interaction $g$, while keeping $g/\delta$ constant.}
	\begin{figure}[h]
		\centering
		\includegraphics{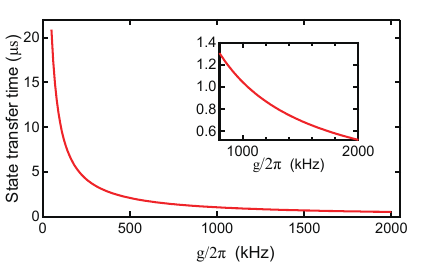}
		\caption[The dependence]{\textbf{The dependence of state transfer time $\tau_\mathrm{ST}$ on the strength of the TMS interaction $g$.} 
			The value of $g/\delta$ is kept to $1/4.66$. The inset is a zoomed-in plot.
		}
		\label{fig:swapvsg}
	\end{figure}
	
	\red{As shown in Fig.~\ref{fig:swapvsg}, while keeping $\delta/g=4.66$, which corresponds to the best fidelity in our current experiments, we plot $\tau_\mathrm{ST}$ as a function of $g$.  From the inset, we find that the state transfer time $\tau_\mathrm{ST}$ falls below $1~\mu$s when $g/2\pi>1.1$ MHz. This is achievable with other kinds of couplers, such as SNAIL~\cite{PRXQuantum.4.020355Ref3} and flux-pumped SQUID~\cite{Lu_NaturecomRef2}.}
	\section{{Comparison between TMS interactions and BS interactions}}
	\begin{figure}[b]
		\centering
		\includegraphics{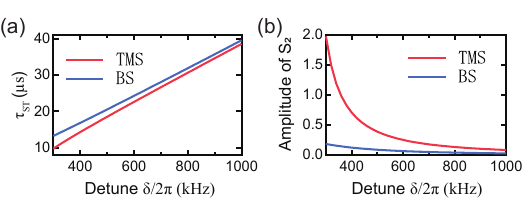}
		\caption[The dependence]{\textbf{The dependence of state transfer time $\tau_\mathrm{ST}$ and oscillation amplitude of photons in mode $S_2$ on $\delta$ for the TMS interactions and BS interactions.} We keep $g/2\pi=80$~kHz in both (a) and (b). 
		}
		\label{fig:TMSvsBS}
	\end{figure}
	Quantum state transfer (QST) between separated modes using BS interactions has been demonstrated in previous work. Here, we compare our QST using TMS interactions with that using BS interactions.
	
	Starting from the Hamiltonian
	\begin{equation}
		\hat H_{}= g_1(\hat a^\dagger_{1} \hat a_{2}+\hat a_{1} \hat a^\dagger_{2})+g_2(\hat a^\dagger_{3} \hat a_{2}+\hat a_{3} \hat a^\dagger_{2})+\delta\hat a^\dagger_{2} \hat a_{2},
		\label{eq:HBS2}
	\end{equation}
	which is formally similar to Eq.~\ref{eq:Hdetune}, we solve the Heisenberg equation of motion. We obtain the state transfer time and the oscillation of population in the intermediate mode $S_2$: 
	\begin{equation} 
		\begin{aligned}
			&\tau_\mathrm{ST}=\pi/(\sqrt{2}\Omega_2-\frac{\delta}{2}),\\
			&n_2(t)=(\frac{g}{\sqrt2\Omega_2})^2[1-\cos(2\sqrt2\Omega_2 t)],\\
		\end{aligned}
		\label{eq:BSswapt}
	\end{equation}
	where $\Omega_2=\sqrt{\frac{\delta^2}{8}+g^2}$. To clearly compare the two methods, we plot these two parameters with respect to $\delta$ under the condition of $g/2\pi=80$~kHz. The results are shown in Fig.~\ref{fig:TMSvsBS}. It can be observed that, for the same detuning $\delta$, state transfer using TMS interactions is faster than that using the BS method. However, this comes at the cost of larger amplitude oscillations of the photon number in the intermediate mode $S_2$. As the detuning $\delta$ increases, the effects of these two methods gradually converge to be similar. In the regime where $g \ll \delta$, our approach is comparable to the protocol using the BS interactions.
	
	
	Another major difference is the pump frequencies required to implement these two interactions. We will consider four-wave mixing elements like transmon and three-wave mixing elements like SNAIL. To achieve the TMS interaction, the frequency matching conditions are
	\begin{equation}
		\begin{aligned}
			\omega_{p_1}\pm\omega_{p_2}&=\omega_{s_{1}}+\omega_{s_2},\\
			\omega_{p}&=\omega_{s_{1}}+\omega_{s_2},\\
		\end{aligned}
		\label{eq:freq}
	\end{equation}
	for four-wave mixing and three-wave mixing, respectively. Whereas to achieve the BS interaction, the conditions required are
	\begin{equation}
		\begin{aligned}
			\omega_{p_1}\pm\omega_{p_2}&=\omega_{s_{1}}-\omega_{s_2},\\
			\omega_{p}&=\omega_{s_{1}}-\omega_{s_2}.
			\label{eq:freq}
		\end{aligned}
	\end{equation}
	Our scheme offers additional possibilities for the pump frequencies when transferring a quantum state. This can provide some convenience in certain scenarios.
	As mentioned above, exchange-free QST allows us to think outside the box of BS interactions. This approach provides an alternative choice for practical experiments, expanding the toolbox of techniques for connecting distant qubits and transferring information across quantum processors. It also offers more flexibility when realizing quantum devices and protocols. To solidify this aspect, we provide three detailed examples as follows. 
	
	The first example is the microwave-optical frequency transduction. Quantum transduction between microwave and optical frequencies is essential for scaling up superconducting quantum information processors. Most experimental schemes rely on three-wave mixing nonlinear interactions, where a coherent drive compensates for the frequency mismatch between the target microwave and optical signals. By tuning the drive frequency, one can realize either BS (red-detuned) or TMS (blue-detuned) interactions. However, optical materials face technical challenges in achieving stable BS interactions with strong red-detuned drives. As explicitly pointed out by Jiang et al. ~\cite{Jiang2020Ref1}, the conversion efficiency differs significantly for red-detuned vs blue-detuned pumps. Our scheme, based on TMS interactions, would enable more efficient realization of QST between microwave and optical frequencies, overcoming the limitations of red-detuned drives.
	
	The second example is in the field of coherent conversion between telecommunication channels. When building large-scale quantum networks, utilizing telecommunication photons to establish remote entanglement between nodes is necessary. Similar to classical networks, optical fibers support many channels (e.g., with 100-GHz spacing) that can greatly enhance communication capacity. Coherent information transfer between these channels will be inevitable. However, the frequency differences are only a few hundreds of GHz or a few THz, making coherent conversion via three-wave mixing challenging due to the lack of drive sources at these frequencies. Four-wave mixing requires two optical drives to stimulate coherent BS interactions, but these drives also create significant frequency sidebands due to parasitic nonlinear processes. Our scheme offers an alternative by driving the system with a wavelength around 780~nm, which is readily available and widely used in optics. This approach circumvents the need for low-frequency or multiple drives, enabling more efficient and flexible implementation of coherent conversion between telecom channels.
	
	The third example is QST in superconducting circuits. In superconducting systems, QST between microwave photons with frequency differences of only a few tens or hundreds of MHz is important. While three-wave mixing devices (e.g., SNAILs) can be driven at such low frequencies, there may be potential prohibitions of low-frequency signals in compact superconducting devices. In the work of Chapman et al.~\cite{PRXQuantum.4.020355Ref3}, in order to provide a strong pump for SNAIL, an additional buffer mode is used, with a frequency close to the pump frequency. However, if we want to use a low-frequency buffer mode to enhance low-frequency signals, the dense high-order modes of the buffer mode could negatively impact other modes in the system, such as the Purcell limit and photon leakage. Additionally, low-frequency RF signals have relatively large wavelengths, which might induce stronger cross-talk between devices. In contrast, our scheme circumvents these issues by driving at high frequencies ($\sim$10 GHz), enabling more efficient and robust QST in superconducting circuits without the limitations associated with low-frequency drives.

\end{document}